\documentclass[8pt]{revtex4}
\pdfoutput=1
\usepackage{amssymb}
\usepackage{latexsym}
\usepackage{epsfig}
\begin{document}

\title{ Emergence and oscillation of cosmic space by joining  M1-branes}

\author{Alireza Sepehri $^{1,2}$}
\email{alireza.sepehri@uk.ac.ir} \affiliation{ $^{1}$Faculty of
Physics, Shahid Bahonar University, P.O. Box 76175, Kerman,
Iran.\\$^{2}$ Research Institute for Astronomy and Astrophysics of
Maragha (RIAAM), P.O. Box 55134-441, Maragha, Iran. }

\author{Farook Rahaman}
\email{rahaman@associates.iucaa.in} \affiliation{Department of
Mathematics, Jadavpur University, Kolkata 700 032, West Bengal,
India.}

\author{Salvatore Capozziello $^{1,2,3,4}$}
\email{capozziello@na.infn.it} \affiliation{$^{1}$ Dipartimento di
Fisica "E. Pancini",  Universit\'a di Napoli "Federico II", I-80126 - Napoli,
Italy. \\
$^{ 2 }$ INFN Sez. di Napoli, Compl. Univ. di Monte S. Angelo,
Edificio G, I-80126 - Napoli, Italy.\\ $^{  3}$ Gran Sasso Science
Institute (INFN), Viale F. Crispi, 7, I-67100, L''Aquila, Italy.\\
$^{4}$Tomsk State Pedagogical University, ul. Kievskaya, 60, 634061 Tomsk, Russia.}

\author{Ahmed Farag Ali}
\email{ahmed.ali@fsu.edu.eg} \affiliation{Deptartment of Physics,
Faculty of Science, Benha University, Benha 13518, Egypt.}

\author{Anirudh Pradhan}
\email{pradhan@associates.iucaa.in} \affiliation{Department of
Mathematics, Institute of Applied Sciences and Humanities, G L A
University, Mathura-281 406, Uttar Pradesh, India}

\begin{abstract}
Recently, it has been proposed by Padmanabhan  that the difference
between the number of degrees of freedom on the boundary surface
and the number of degrees of freedom in a bulk region leads to the
expansion of the universe. Now, a natural question arises; how
this model could explain the oscillation of universe between
contraction and expansion branches? We try to address this issue
in the framework of BIonic system. In this model, $M0$-branes join to
each other and give rise to  a pair of $M1$-anti-$M1$-branes. The fields
which live on these branes play the roles of  massive gravitons
that cause the emergence of a wormhole between them and
formation of a BIon system. This wormhole dissolves into M1-branes and
causes a divergence between the number of degrees of freedom on
the boundary surface of $M1$ and the bulk leading to an expansion
of $M1$-branes. When $M1$-branes become close to each other, the
square energy of their system becomes negative and some tachyonic
states  emerge. To removes these states, $M1$-branes compact,
the sign of compacted gravity changes, causing the arising of
anti-gravity: in this case,   branes get away from each other. By articulating
$M1$-BIons, an M3-brane and an anti-$M3$-brane are created
and connected by three wormholes  forming an $M3$-BIon. This new
system behaves like the initial system and by closing branes to
each other, they compact and, by getting away from each other, they
open. Our universe is located on one of these M3-branes and, by
compacting $M3$-brane, it contracts and, by opening it, it
expands.\\\\


 \end{abstract}
 \date{\today}


 \maketitle

\section{Introduction}
The origin of the universe expansion has been described recently
by Padmanabhan \cite{q1}. It has been proposed that the expansion
of the universe happens as a result of a deviation between the
surface degrees of freedom on the holographic horizon and the bulk
degrees of freedom \cite{q1}. To date, several papers investigated
this interesting proposal and its implications for cosmology
\cite{q2,q3,q4,q5,q6,q7,q8}. For example, the Padmanabhan
proposal has been used to deduce the Friedmann equations of an ($n
+ 1$)-dimensional Friedmann-Robertson-Walker (FRW) universe in the
framework of general relativity, Gauss-Bonnet gravity and Lovelock
gravity  \cite{q2}. In another case, the proposal has been
extended to brane cosmology, scalar-tensor cosmology and $f(R)$
gravity \cite{q3}. In another scenario, with the help of
Padmanabhan's proposal, it has been derived the Friedmann
equations of universe  in higher dimensional space-time in
different gravities like Gauss-Bonnet and Lovelock gravity with
general spacial curvature  \cite{q4}. In another investigation,
the Padmanabhan idea has been generalized to the non-flat
universe corresponding to the spatial curvature parameter $k = \pm 1$ \cite{q5,q6}. Besides, in
\cite{q7},  the Padmanabhan proposal has been investigated in the
context of Generalized Uncertainty Principle (GUP). And in a
recent work, the Padmanabhan model has been considered in BIonic
system and it has been argued that the difference between the degrees
of freedom inside and outside the universe is due to the
evolutions of branes in extra dimensions \cite{q8}. In general, the BIon is a
configuration of two branes which are connected by a wormhole
\cite{q9,q10,q11,q12}. \\

On the other hand, recent investigations show that the universe
may oscillate between contraction and expansion branches
\cite{q13,q14}. A naturally arising question is whether the
Padmanabhan proposal could explain the universe oscillation. We
try to answer this question in the framework of a BIonic system. In previous
studies, it has been argued that the Big Bang may be removed in
string theory and replaced by $N$ fundamental strings \cite{q12}.
In this model, first, $N$ fundamental strings transit to $N$ pairs
of $M0$-brane and anti-$M0$-brane. Then, these branes glue to each
other and build up a BIonic system which is a configuration of
$M3$-brane, and anti-$M3$-brane in addition to a wormhole. Our
universe is located on one of these $M3$-branes and interacts with
other universe via the wormhole \cite{q12}.

In this paper, we will
extend those calculations and show that by joining $M0$-branes, a
pair of $M1$-anti-$M1$-branes could be constructed. At that
stage, two types of fields are produced and interact with branes.
One type plays the role of scalar field  in transverse dimensions
and another one appears as graviton fields on the $M1$-branes.
These gravitons lead to the emergence of a wormhole between branes
and hence the formation of a BIon system. The evolution of BIon leads to
the difference between the number of degrees of freedom on the
boundary surface of M1 and the bulk and this difference is the
main cause of the expansion of M1-branes in the Padmanabhan
picture.  When $M1$-branes approach to each other, the square
energy of branes system becomes negatives and the system
transits to the tachyon phase. To remove these tachyon states,
$M1$-branes become compact and gravity turns  to be anti-gravity.
In that conditions, branes get away from each other and begin to
be opened. These BIons glue  each other and form a bigger BIon
which includes $M3$-brane and anti-$M3$-brane in addition to
three wormholes connecting them. The $M3$-branes oscillate between
compact and open branches as due to the oscillation of initial
$M1$-branes. Our universe is placed on one of these $M3$-branes.
By compacting the $M3$-branes, it contracts and by
opening $M3$-branes, it expands. \\

The outline of the paper is the  following.  In section \ref{o1}, we
consider the formation and the expansion of $M1$-branes. We also study
the process of formation of $M3$ from M1-branes and obtain the
difference between the number of degrees of freedom of the universe in
terms of  BIon evolution. In section \ref{o2}, we discuss how,
by compacting branes, gravity turns  to  anti-gravity and the
contraction of the branches begins. The last section is devoted to discussion and conclusions.

\section{ Cosmic expansion in Padmanabhan model
}\label{o1}

In previous studies, it has been shown that by joining $M0$-branes
to each other, a pair of $M1$-anti-$M1$-branes can be formed
\cite{q12}. The fields on these branes play the role of graviton
and cause the formation of a wormhole between these branes. These
graviton fields are the main cause of difference between the
number of degrees of freedom of brane and bulk and hence causing
an expansion. By closing $M1$-branes, they bend, compact and
gravity turns to be anti-gravity. By gluing $M1$-branes,
$M3$-branes are formed which our universe is located on one of
them. These branes expand and compact like the initial M1-branes
and this fact leads to an oscillation of our universe  between
expansion and contraction branches. The action of $M1$ can be
written as \cite{q12,q15,q16,q17,q18,q19,qq19,q20,q21,q22,q23}:

\begin{eqnarray}
 S = - T_{M1}\int  d^{2}\sigma ~ STr \sqrt{-det(P_{abc}[ E_{mnl}
+E_{mij}(Q^{-1}-\delta)^{ijk}E_{kln}]+\nonumber \lambda
F_{abc})det(Q^{i}_{j,k})}~~ \label{m1}
\end{eqnarray}
where

\begin{eqnarray}
 E_{mnl}^{\alpha,\beta,\gamma} &=& G_{mnl}^{\alpha,\beta,\gamma} + B_{mnl}^{\alpha,\beta,\gamma}, \\
 Q^{i}_{j,k} &=& \delta^{i}_{j,k} + i\lambda[X^{j}_{\alpha}T^{\alpha},X^{k}_{\beta}T^{\beta},X^{k'}_{\gamma}T^{\gamma}]
 E_{k'jl}^{\alpha,\beta,\gamma},\nonumber\\F_{abc}&=&\partial_{a} A_{bc}-\partial_{b} A_{ca}+\partial_{c}
A_{ab}. \label{m2}
\end{eqnarray}
Here $X^{M}=X^{M}_{\alpha}T^{\alpha}$, $A_{ab}$ is $2$-form gauge
field,

\begin{eqnarray}
 &&[T^{\alpha}, T^{\beta}, T^{\gamma}]=
 f^{\alpha \beta \gamma}_{\eta}T^{\eta} \nonumber \\&& [X^{M},X^{N},X^{L}]=
 [X^{M}_{\alpha}T^{\alpha},X^{N}_{\beta}T^{\beta},X^{L}_{\gamma}T^{\gamma}]
\label{m3}
\end{eqnarray}
where $\lambda=2\pi l_{s}^{2}$,
$G_{mnl}=g_{mn}\delta^{n'}_{n,l}+\partial_{m}X^{i}\partial_{n'}X^{i}\sum_{j}
(X^{j})^{2}\delta^{n'}_{n,l} +
\partial_{n'}\partial_{m}X^{i}\partial_{m}\partial_{n'}X^{i}\delta^{n'}_{n,l} $ and $X^{i}$ are scalar fields
of mass dimension. Here $a,b=0,1,...,p$ are the world-volume
indices of the $Mp$-branes, $i,j,k = p+1,...,9$ are indices of the
transverse space, and m,n are the eleven-dimensional spacetime
indices. Also, $T_{Mp}=1/\left(g_{s}(2\pi)^{p}l_{s}^{p+1}\right)$
is the tension of Mp-brane, $l_{s}$ is the string length and
$g_{s}$ is the string coupling. In previous studies, it has been
shown that this action can be obtained by summing over the actions
of $p M0$-branes which is given by \cite{q12}:

\begin{eqnarray}
S_{M0} = T_{M0}\int dt Tr\left( \Sigma_{M,N,L=0}^{10}
\langle[X^{M},X^{N},X^{L}],[X^{M},X^{N},X^{L}]\rangle\right)
\label{m4}
\end{eqnarray}
To obtain the action (\ref{m1}) from the action of M0, we should
use of following mappings
\cite{q12,q15,q16,q17,q18,q19,qq19,q20,q21,q22,q23}:

\begin{eqnarray}
&&\langle[X^{a},X^{i},X^{j}],[X^{a},X^{i},X^{j}]\rangle=
\frac{1}{2}\varepsilon^{abc}\varepsilon^{abd}(\partial_{c}X^{i}_{\alpha})(\partial_{d}X^{i}_{\beta})
\langle(T^{\alpha},T^{\beta})\rangle \sum_{j} (X^{j})^{2} =
 \frac{1}{2}\langle \partial_{a}X^{i},\partial_{a}X^{i}\rangle \sum_{j} (X^{j})^{2}\nonumber \\
&&\nonumber \\
&&\nonumber \\
&& \langle[X^{a},X^{b},X^{j}],[X^{a'},X^{b'},X^{j}]\rangle=
\frac{1}{2}\varepsilon^{abc}\varepsilon^{a'b'c}(\partial_{a}\partial_{b}X^{i}_{\alpha})(\partial_{a'}
\partial_{b'}X^{i}_{\beta})\langle(T^{\alpha},T^{\beta})\rangle
 =
 \frac{1}{2}\langle \partial_{a}\partial_{b}X^{i}_{\alpha},\partial_{a'}\partial_{b'}X^{i}_{\alpha}\rangle \nonumber \\
&&\langle[X^{a},X^{b},X^{c}],[X^{a},X^{b},X^{c}]\rangle=
(F^{abc}_{\alpha\beta\gamma})(F^{abc}_{\alpha\beta\eta})\left(\langle[T^{\alpha},T^{\beta},T^{\gamma}],[T^{\alpha},T^{\beta},T^{\eta}]
\rangle\right)=\nonumber \\
&&
(F^{abc}_{\alpha\beta\gamma})(F^{abc}_{\alpha\beta\eta})f^{\alpha
\beta \gamma}_{\sigma}h^{\sigma \kappa}f^{\alpha \beta
\eta}_{\kappa} \langle T^{\gamma},T^{\eta}\rangle =
(F^{abc}_{\alpha\beta\gamma})(F^{abc}_{\alpha\beta\eta})\delta^{\kappa
\sigma} \langle T^{\gamma},T^{\eta}\rangle= \langle
F^{abc},F^{abc}\rangle \nonumber \\
&&\nonumber \\
&& i,j=p+1,..,10\quad a,b=0,1,...p\quad m,n=0,..,10~~ \label{m5}
\end{eqnarray}
To obtain a similarity between branes and our real world, we
assume that two form fields play the role of gravitons and obtain
following results:

\begin{eqnarray}
&&A^{ab}=g^{ab}=h^{ab}+\eta^{ab} ~ ~  and  ~ ~
a,b,c=\mu,\nu,\lambda \Rightarrow \nonumber\\&&
F_{abc}=\partial_{a} A_{bc}-\partial_{b} A_{ca}+\partial_{c}
A_{ab}=2(\partial_{\mu}g_{\nu\lambda}+\partial_{\nu}g_{\mu\lambda}-\partial_{\lambda}g_{\mu\nu})=
2\Gamma_{\mu\nu\lambda}\nonumber\\&&\nonumber\\&&\langle
F^{\rho}\smallskip_{\sigma\lambda},F^{\lambda}\smallskip_{\mu\nu}\rangle=
\langle[X^{\rho},X_{\sigma},X_{\lambda}],[X^{\lambda},X_{\mu},X_{\nu}]\rangle=\nonumber\\&&
[X_{\nu},[X^{\rho},X_{\sigma},X_{\mu}]]-[X_{\mu},[X^{\rho},X_{\sigma},X_{\nu}]]+[X^{\rho},X_{\lambda},X_{\nu}]
[X^{\lambda},X_{\sigma},X_{\mu}]
-[X^{\rho},X_{\lambda},X_{\mu}][X^{\lambda},X_{\sigma},X_{\nu}]=\nonumber\\&&\partial_{\nu}
\Gamma^{\rho}_{\sigma\mu}-\partial_{\mu}\Gamma^{\rho}_{\sigma\nu}+\Gamma^{\rho}_{\lambda\nu}
\Gamma^{\lambda}_{\sigma\mu}-\Gamma^{\rho}_{\lambda\mu}\Gamma^{\lambda}_{\sigma\nu}
=R^{\rho}_{\sigma\mu\nu}\nonumber\\&&\nonumber\\&&\nonumber
\end{eqnarray}
and
\begin{eqnarray}
&&\kappa^{\mu}_{\nu}=\delta^{\mu}_{\nu}-\sqrt{\delta^{\mu}_{\nu}-H^{\mu}_{\nu}}\nonumber\\
&&H_{\mu\nu}=
g_{\mu\nu}-\eta_{mn}\partial_{\mu}X^{m}\partial_{\nu}X^{n}\nonumber\\
&&H_{\mu\nu}=h_{\mu\nu}+2\Pi_{\mu\nu}-\eta^{\alpha\beta}\Pi_{\mu\alpha}\Pi_{\beta\nu}\nonumber\\
&&X^{m}=x^{m}-\eta^{m\mu}\partial_{\mu}\pi \nonumber\\
&&\Pi_{\mu\nu}=\partial_{\mu}\partial_{\nu}\pi \label{m6}
\end{eqnarray}
where $\pi$ is the scalar mode  and $h^{ab}$ is the tensor mode of
graviton. As can be seen from above equations, non-commutative
relations between two form fields produce the exact form of
curvature tensor. Also, when scalars are attached to branes, their
index changes from $i,j\rightarrow \mu\nu$ and they transit to
graviton mode. Previously, it has been shown that there are direct
relations between $\kappa$ and curvature scalars ($R$)
\cite{q24,q25,q26}:

\begin{eqnarray}
&&\delta^{\rho\sigma}_{\mu\nu}\kappa^{\mu}_{\rho}\kappa^{\nu}_{\sigma}=R
\label{m7}
\end{eqnarray}
Thus, gravity  can be easily obtained from the non-commutative
relations  in $M$-theory. At this stage, we can derive the
explicit form of the relevant action of $M1$ in equation
(\ref{m1}) in terms of gravity terms. We can write:
\begin{eqnarray}
&&
\det(Z)=\delta^{a_{1},a_{2}...a_{n}}_{b_{1}b_{2}....b_{n}}Z^{b_{1}}_{a_{1}}...Z^{b_{n}}_{a_{n}}
\quad a,b,c=\mu,\nu,\lambda \nonumber\\\nonumber\\
&&Z_{abc}=P_{abc}[ E_{mnl} +E_{mij}(Q^{-1}-\delta)^{ijk}E_{kln}]+
\lambda F_{abc}\nonumber\\&&\nonumber\\
&&\det(Z)=\det\left(P_{abc}[ E_{mnl}
+E_{mij}(Q^{-1}-\delta)^{ijk}E_{kln}]\right)+\lambda^{2}\det(F)\label{m8}\end{eqnarray}
This equation helps us to derive the relevant terms of determinant
in action (\ref{m1}) separately. Applying relations in equation
(\ref{m7}) in determinants( \ref{m8}), we obtain

\begin{eqnarray}
\det(F)=\delta_{\rho\sigma}^{\mu\nu}\langle
F^{\rho\sigma}\smallskip_{\lambda},F^{\lambda}\smallskip_{\mu\nu}\rangle
=\delta_{\rho\sigma}^{\mu\nu}R^{\rho\sigma}_{\mu\nu}\label{m9}\end{eqnarray}

\begin{eqnarray}
&& \det(P_{abc}[ E_{mnl}
+E_{mij}(Q^{-1}-\delta)^{ijk}E_{kln}])=\nonumber\\&&\delta_{\rho\sigma}^{\mu\nu}
[(g^{\mu}_{\rho}g^{\nu}_{\sigma}+ g^{\nu}_{\sigma}\langle
\partial^{\mu}X^{i},\partial_{\rho}X^{j}\rangle \sum (X^{i})^{2}+\langle
\partial^{\mu}\partial^{\nu}X^{i},\partial_{\sigma}\partial_{\rho}X^{j}\rangle+..)-
\nonumber\\&&\frac{(g^{\mu}_{\rho}g^{\nu}_{\sigma}+
g^{\nu}_{\sigma}\langle
\partial^{\mu}X^{i},\partial_{\rho}X^{j}\rangle \sum (X^{i})^{2}+\langle
\partial^{\mu}\partial^{\nu}X^{i},\partial_{\sigma}\partial_{\rho}X^{j}\rangle+...)}{[(\lambda)^{2}
\det([X^{j}_{\alpha}T^{\alpha},X^{k}_{\beta}T^{\beta},X^{k'}_{\gamma}T^{\gamma}])]}]=
\nonumber\\&&\delta_{\rho\sigma}^{\mu\nu}[\kappa_{\mu}^{\rho}\kappa_{\nu}^{\sigma}
\sum (X^{i})^{2}+
(\partial_{\lambda}\kappa_{\mu}^{\rho}\partial^{\lambda}\kappa_{\nu}^{\sigma})]
(1-\frac{1}{[(\lambda)^{2}\det([X^{j}_{\alpha}T^{\alpha},X^{k}_{\beta}T^{\beta},X^{k'}_{\gamma}T^{\gamma}])]})=
\nonumber\\&&\delta_{\rho\sigma}^{\mu\nu}[\kappa_{\mu}^{\rho}\kappa_{\nu}^{\sigma}
\sum (X^{i})^{2}+
(\partial_{\lambda}\kappa_{\mu}^{\rho}\partial^{\lambda}\kappa_{\nu}^{\sigma})]\left(1-\frac{1}{m_{g}^{2}}\right)\label{m10}
\end{eqnarray}
where
$m_{g}^{2}=[(\lambda)^{2}\det([X^{j}_{\alpha}T^{\alpha},X^{k}_{\beta}T^{\beta},X^{k'}_{\gamma}T^{\gamma}])]
$ is the square of graviton mass. It is clear that the graviton
mass depends on the scalars which interact with branes. This is
because  when scalars collide with branes, their index changes
and they transit to graviton.  With this definition, we can
calculate another term of the determinant:

\begin{eqnarray}
 &&\det(Q)\sim
(i)^{2}(\lambda)^{2}\det([X^{j}_{\alpha}T^{\alpha},X^{k}_{\beta}T^{\beta},X^{k'}_{\gamma}T^{\gamma}])\det(E)\sim
\nonumber\\&&
-[(\lambda)^{2}\det([X^{j}_{\alpha}T^{\alpha},X^{k}_{\beta}T^{\beta},X^{k'}_{\gamma}T^{\gamma}])]\det(g)=m_{g}^{2}\det(g)
\label{m11}
\end{eqnarray}

By inserting equations (\ref{m7}) ,(\ref{m9}), (\ref{m10}) and
(\ref{m11}) into the action (\ref{m1}) and replacing $\sum
(X^{i})^{2}\rightarrow F(X)$, we get:

\begin{eqnarray}
&& S_{M1} =  -T_{M1} \int d^{2}\sigma
\Bigl[\sqrt{-g}\Big(\delta^{\rho\sigma}_{\mu\nu}[\kappa_{\mu}^{\rho}\kappa_{\nu}^{\sigma}
\sum (X^{i})^{2}+
(\partial_{\lambda}\kappa_{\mu}^{\rho}\partial^{\lambda}\kappa_{\nu}^{\sigma})]-
m_{g}^{2}\delta^{\rho\sigma}_{\mu\nu}\left(R_{\rho\sigma}^{\mu\nu}+[\kappa_{\mu}^{\rho}\kappa_{\nu}^{\sigma}
\sum (X^{i})^{2}+
(\partial_{\lambda}\kappa_{\mu}^{\rho}\partial^{\lambda}\kappa_{\nu}^{\sigma})]\right)\Big)\Bigr]=\nonumber\\&&
-T_{M1} \int d^{2}\sigma \Bigl[\sqrt{-g}\Big(F(X)R-
m_{g}^{2}\delta^{\rho\sigma}_{\mu\nu}R_{\rho\sigma}^{\mu\nu}-m_{g}^{2}F(X)R+\delta_{\rho\sigma}^{\mu\nu}(1-m_{g}^{2})
(\partial_{\lambda}\kappa_{\mu}^{\rho}\partial^{\lambda}\kappa_{\nu}^{\sigma})\Big)\Bigr]
\label{m12}
\end{eqnarray}
Obviously,  first order  terms in nonlinear theories, like
Lovelock and massive gravity,  are present in this action. This
means that there is a direct relation between $M$-theory and effective
theories  of gravity. According to these calculations, there are two types of
modes for gravitons. Scalar modes which  are produced by attaching
scalars to branes and tensor modes which are produced in the
process of formation of M1 from M0-branes (see also \cite{rept,bogdanos}).

Using the equation
(\ref{m1})and assuming the separation distance between two M1 be
$l_{d}$ and the length of each M1 be $l_{1}$, we can obtain the
relevant action for the interaction of an M1 with an
anti-$M1$-brane:

\begin{eqnarray}
&&  A^{ab}\rightarrow l_{1} \quad X^{2}\rightarrow l_{d} \quad X^{0}=t \quad X^{i}=0, i\neq 0,2\nonumber\\
&& S=-T_{M1}\int d^{2}\sigma \sqrt{l_{d}^{2}+l_{d}^{5}}
\sqrt{\Bigl[l_{1}^{2}+(l'_{d})^{2}+(l''_{d})^{2}(1+l_{d}^{2})^{-1}\Bigr]\Bigl[1-\frac{1}{l_{d}^{3}}\Bigr]-(l'_{1})^{2}(l_{d})^{2}}=
\nonumber\\
&&-T_{M1}\int d^{2}\sigma V(l_{d})
\sqrt{D_{l_{d},l_{1}}}\nonumber\\&&
 V(l_{d})= \sqrt{l_{d}^{2}+l_{d}^{5}}\nonumber\\&&
 D_{l_{d},l_{1}}=\Bigl[l_{1}^{2}+(l'_{d})^{2}+(l''_{d})^{2}(1+l_{d}^{2})^{-1}\Bigr]\Bigl[1-\frac{1}{l_{d}^{3}}\Bigr]-
 (l'_{1})^{2}(l_{d})^{2}\label{m13}
\end{eqnarray}
where the prime denotes the derivative respect to time. The
equations of motion obtained from this action are:

\begin{eqnarray}
&&
\Bigl(\frac{(l'_{d})\left[1-\frac{1}{l_{d}^{3}}\right]}{\sqrt{D_{l_{d},l_{1}}}}\Bigr)'=\frac{1}{\sqrt{D_{l_{d},l_{1}}}}
\Bigl(2l_{d}^{-4}[(l'_{d})^{2}+(l''_{d})^{2}(1+l_{d}^{2})^{-1}](l'_{d})
-2(l'_{1})^{2}(l_{d})(l'_{d})+\frac{V'}{V}\Biggl[D_{l_{d},l_{1}}-(l'_{d})\Biggl[1-\frac{1}{l_{d}^{3}}\Biggr]\Biggr]\Bigr)\nonumber\\&&
\Big(\frac{2(l'_{1})(l_{d})^{2}}{\sqrt{D_{l_{d},l_{1}}}}\Big)'=
\frac{1}{\sqrt{D_{l_{d},l_{1}}}}\Bigl(l_{1}\left[1-\frac{1}{l_{d}^{3}}\right]-\frac{V'}{V}\left[l'_{1}l_{d}^{2}\right]\Bigr)\label{m14}
\end{eqnarray}
Solving these equations simultaneously, we  obtain the
approximate form of $l_{d}$ and $l_{1}$ in terms of time:

\begin{eqnarray}
&& l_{1}\sim
\frac{l_{1,+}}{(t_{s}-t)^{2}}(\frac{t_{s}}{(t_{s}-t)}-1)e^{-l_{0}(t_{s}-t)}
\nonumber\\&& l_{d}\sim
\frac{l_{d,0}}{t_{s}^{1/2}}(t_{s}-t)^{1/2}e^{\frac{-t}{t_{s}}}\label{m15}
\end{eqnarray}

where $t_{s}$ is time of collision between $M1$-branes and
$l_{d,0}$ is the maximum distance between two $M1$-branes. To be
sure that these solutions are true, specially near the point that
branes collide to each other, we consider their correctness when
branes are closed to each other ($l_{d}\sim 0$). In this case, the
size of two branes is very big ($l_{1}\sim \infty$) and the
velocity of their motion and rate of their growth is large
($l'_{d}\sim l'_{1} \sim \infty$).For this state, equations in
(\ref{m14}) reduce to following equations:

\begin{eqnarray}
&&   l_{d}\sim 0 \quad l'_{d}\sim \infty \nonumber\\&&
\Rightarrow\frac{V'}{V}D_{l_{d},l_{1}}\sim
\frac{l'_{d}(2l_{d}+5l_{d}^{4})}{l_{d}^{2}+l_{d}^{5}}\Bigl(\Bigl[l_{1}^{2}+(l'_{d})^{2}+(l''_{d})^{2}(1+l_{d}^{2})^{-1}\Bigr]\Bigl[1-\frac{1}{l_{d}^{3}}\Bigr]-
 (l'_{1})^{2}(l_{d})^{2}\Bigr)\sim\nonumber\\&&2l_{d}^{-4}[(l'_{d})^{2}+(l''_{d})^{2}(1+l_{d}^{2})^{-1}](l'_{d})
-2(l'_{1})^{2}(l_{d})(l'_{d})
\nonumber\\&&\nonumber\\&&\nonumber\\&&
\Rightarrow\Bigl(\frac{(l'_{d})\left[1-\frac{1}{l_{d}^{3}}\right]}{\sqrt{D_{l_{d},l_{1}}}}\Bigr)'\sim\frac{1}{\sqrt{D_{l_{d},l_{1}}}}
\Bigl((l'_{d})\left[1-\frac{1}{l_{d}^{3}}\right]\Bigr)\nonumber\\&&\Rightarrow
X=\frac{1}{\sqrt{D_{l_{d},l_{1}}}}
\Bigl((l'_{d})\left[1-\frac{1}{l_{d}^{3}}\right]\Bigr)\rightarrow
\frac{d X}{dt}=\lambda X\rightarrow X=e^{\lambda
t}\nonumber\\&&Y=\frac{1}{X}\sim
\frac{(l''_{d})^{2}(1+l_{d}^{2})^{-1}}{(l'_{d})^{2}\left[1-\frac{1}{l_{d}^{3}}\right]^{2}}\nonumber\\&&\Rightarrow
l_{d}\sim (B-t)^{\frac{1}{2}}e^{-\frac{\lambda t}{2}}
\nonumber\\&& l_{d}(t=t_{s})=0\rightarrow B=t_{s} \quad
\lambda=\frac{2}{t_{s}} \nonumber\\&&\nonumber\\&&\nonumber\\&&
\Big(\frac{(l'_{1})(l_{d})^{2}}{\sqrt{D_{l_{d},l_{1}}}}\Big)'\simeq
\frac{1}{\sqrt{D_{l_{d},l_{1}}}}\Bigl(l'_{1}l_{d}^{2}\Bigr)\nonumber\\&&l_{d}\sim
(t_{s}-t)^{\frac{1}{2}}e^{-\frac{ t}{t_{s}}}  \Rightarrow
D_{l_{d},l_{1}}^{\frac{1}{2}}\sim (t-t_{s})^{-3}
\nonumber\\&&\Rightarrow \Big(l'_{1} (t_{s}-t)^{4}\Big)'\simeq
\Big(l'_{1} (t_{s}-t)^{4}\Big)\nonumber\\&&\Rightarrow l_{1}\sim
\frac{1}{(t_{s}-t)^{2}}(\frac{t_{s}}{(t_{s}-t)}-1)e^{-l_{0}(t_{s}-t)}
\label{1m14}
\end{eqnarray}

This equation shows that by passing time, two $M1$-branes move
towards each other and $l_{d}$ decreases while the size of $M1$
increases. We can show that a wormhole is formed between these
$M1$-branes that ,by dissolving in them,  causes to their growth.
Before discussing this subject, we will construct Mp-branes from
gluing M1-branes. To this end, by  equation (\ref{m6}), we use the
following replacements in the action of the branes.

\begin{eqnarray}
&&i,j=a,b\Rightarrow
\langle[X^{i},X^{j},X^{k}],[X^{i},X^{j},X^{k}]\rangle \Rightarrow
\langle[X^{a},X^{j},X^{i}],[X^{a},X^{j},X^{i}]\rangle=
\frac{1}{2}\langle \partial_{a}X^{i},
\partial_{a}X^{i}\rangle \sum X_{j}^{2} \nonumber \\
&&\nonumber \\
&&i,j=a,b\Rightarrow
\langle[X^{i},X^{j},X^{k}],[X^{i},X^{j},X^{k}]\rangle \Rightarrow
\langle[X^{a},X^{b},X^{i}],[X^{a},X^{b},X^{i}]\rangle=
\frac{1}{2}\langle \partial_{b}
\partial_{a}X^{i},\partial_{b}\partial_{a}X^{i}\rangle \nonumber \\
&&\nonumber \\
&&i,j,k=a,b,c\Rightarrow
\langle[X^{i},X^{j},X^{k}],[X^{i},X^{j},X^{k}]\rangle \Rightarrow
\langle[X^{a},X^{b},X^{c}],[X^{a},X^{b},X^{c}]\rangle=  \langle
F^{abc},F^{abc}\rangle \nonumber \\
&&\nonumber \\
&& Q^{i}_{j,k} = \delta^{i}_{j,k}
    + i\lambda[X^{j}_{\alpha}T^{\alpha},X^{k}_{\beta}T^{\beta},X^{k'}_{\gamma}T^{\gamma}]
    E_{k'jl}^{\alpha,\beta,\gamma}\Rightarrow \nonumber \\
&& Q^{i}_{j,k} = \delta^{i}_{j,k}
    + i\Big(\langle[X^{a},X^{j},X^{i}],[X^{a},X^{j},X^{i}]\rangle+
    \langle[X^{a},X^{b},X^{i}],[X^{a},X^{b},X^{i}]\rangle+\langle[X^{a},X^{b},X^{c}],[X^{a},X^{b},X^{c}]\rangle\Big)
    E_{k'jl}^{\alpha,\beta,\gamma}=\nonumber \\
&&\delta^{i}_{j,k}
    + i\Big(\frac{1}{2}\langle \partial_{b} \partial_{a}X^{i},\partial_{b}\partial_{a}X^{i}\rangle+
    \frac{1}{2}\langle \partial_{a}X^{i},\partial_{a}X^{i}\rangle \sum X_{j}^{2}+\langle
F^{abc},F^{abc}\rangle\Big)E_{k'jl}^{\alpha,\beta,\gamma} \nonumber \\
&&\nonumber \\
&& T_{M1}\int d^{2}\sigma \Rightarrow T_{Mp}\int d^{p}\sigma
\label{m16}
\end{eqnarray}
Applying these relations in equation (\ref{m1}), we obtain:

\begin{eqnarray}
&& S=-T_{Mp} \int d^{p}\sigma \sqrt{-det ( O+2\pi
l_{s}^{2}G(F))}\nonumber\\&&
G=\sum_{n=0,..p}\frac{1}{n!}(-\frac{F}{\beta^{2}})^{n}
\nonumber\\&&\nonumber\\&&O=\frac{1}{p}\sum_{n}(p-n)!\frac{Y^{n}}{n!}
\nonumber\\&&\nonumber\\&& F=\langle F^{abc},F^{abc}\rangle\quad
Y=\langle
\partial_{a}X^{i},\partial^{a}X^{i}\rangle\sum (X^{j})^{2}+\langle
\partial_{a}\partial_{b}X^{i},\partial^{a}\partial^{b}X^{i}\rangle \quad \beta=\frac{1}{2\pi
l_{s}^{2}} \label{m17}
\end{eqnarray}
where the nonlinear field ($G$) has been introduced in
\cite{q27,q28,q29}. Now, we can show that this action can be
reproduced by multiplying the terms of relevant actions of $p$
$M1$'s.
 For simplicity, we choose
 $X^{1}=\sigma$ and $X^{4}=z$, $\sum (X_{i})^{2}\rightarrow F(z)$ where $z$ is the transverse direction
between branes. Using the action in (\ref{m17}), the  Lagrangian
for $Mp$-brane can be written as

\begin{eqnarray}
&&\L=- T_{Mp} \int d\sigma \sigma^{p}\sqrt{(1+
z'^{2}F(z)+z''^{2})^{p}+(2\pi l_{s}^{2})^{2}G(F)}\label{m18}
\end{eqnarray}
where ($'$) denotes the derivative respect to $\sigma$ and $z'$
and $z''$ are the velocity and acceleration of branes in
transverse dimension. To derive the Hamiltonian, we must obtain
the canonical momentum density for graviton. For simplicity, we
will use of the method in \cite{q30} and \cite{q31} and assume
that $F_{001}\neq 0$ and other components of F are zero. We get:

\begin{eqnarray}
&&\Pi=\frac{\delta \L}{\delta
\partial_{t}A^{01}}=\frac{\sum_{n=0}^{p}\frac{n}{n!}(-\frac{F}{\beta^{2}})^{n-1}F_{001}}{\sqrt{(1+z'^{2}F(z)+z''^{2})^{p}+
(2\pi l_{s}^{2})^{2}G(F)}} \label{m19}.
\end{eqnarray}
Thus the Hamiltonian can be written as:
\begin{eqnarray}
&&H= T_{Mp}\int d\sigma
\sigma^{p}\Pi\partial_{t}A^{01}-\L=4\pi\int d\sigma [
\sigma^{p}\Pi(\partial_{t}A^{01}-\partial_{\sigma}A^{00})-\partial_{\sigma}(\sigma^{2}\Pi)A_{00}]-\L
\label{m20}
\end{eqnarray}
where we use the integration by parts and applied the term
proportional to $\partial_{\sigma}A^{01}$. Using the constraint
($\partial_{\sigma}(\sigma^{p}\Pi)=0$), we obtain \cite{q30}:

\begin{eqnarray}
&& \Pi=\frac{k}{4\pi \sigma^{p}} \label{m21}
\end{eqnarray}
where $k$ is a constant. Replacing $\Pi$ from the above equation into
equation (\ref{m20}) gives the following Hamiltonian:

\begin{eqnarray}
&&H_{1}= T_{Mp}\int d\sigma \sigma^{p}
\sqrt{(1+z'^{2}F(z)+z''^{2})^{p}+(2\pi
l_{s}^{2})^{2}\sum_{n=0}^{p}\frac{n}{n!}(-\frac{F}{\beta^{2}})^{n}}F_{1}\nonumber\\&&F_{1}=\sqrt{1+\frac{k^{2}}{\sigma^{2p}}}
\label{m22}
\end{eqnarray}
For obtaining the explicit form of the wormhole between branes, we
need a Hamiltonian which can be expressed in terms of separation
distance between branes. To this end, the Lagrangian can be
redefined as:
\begin{eqnarray}
&&\L=- T_{Mp} \int d\sigma
\sigma^{p}\sqrt{(1+z'^{2}F(z)+z''^{2})^{p}+(2\pi
l_{s}^{2})^{2}\sum_{n=0}^{p}\frac{n}{n!}(-\frac{F}{\beta^{2}})^{n}}F_{1}\label{m23}
\end{eqnarray}
With the help of this Lagrangian, we repeat our previous
calculations. We can obtain :
\begin{eqnarray}
&&\Pi=\frac{\delta \L}{\delta
\partial_{t}A^{01}}=\frac{\sum_{n}^{p}\frac{n(n-1)}{n!}(-\frac{F}{\beta^{2}})^{n-1}F_{001}}{\sqrt{(1+z'^{2}F(z)+z''^{2})^{p}+(2\pi
l_{s}^{2})^{2}\sum_{n=0}\frac{n}{n!}(-\frac{F}{\beta^{2}})^{n}}}
\label{m24}
\end{eqnarray}
Therefore the new Hamiltonian can be constructed as:
\begin{eqnarray}
&&H_{2}= T_{Mp}\int d\sigma
\sigma^{p}F_{1}\Pi\partial_{t}A^{01}-\L=\int d\sigma [
\sigma^{p}F_{1}\Pi(\partial_{t}A^{01}-\partial_{\sigma}A^{00})-\partial_{\sigma}(F_{1}\sigma^{p}\Pi)A_{00}]-\L
\label{m25}
\end{eqnarray}
where like the previous step, we have used in the second step an
integration in parts for the term proportional to
$\partial_{\sigma}A^{01}$ like the method in \cite{q30}. Imposing
the constraint ($\partial_{\sigma}(\sigma^{p}\Pi)=0$), we obtain:
\begin{eqnarray}
&& \Pi=\frac{k}{4F_{1}\pi \sigma^{p}} \label{m26}
\end{eqnarray}
By replacing the momentum in equation (\ref{m26}) into equation
(\ref{m25}) we derive the following Hamiltonian:
\begin{eqnarray}
&&H_{2}= T_{Mp}\int d\sigma
\sigma^{p}\sqrt{(1+z'^{2}F(z)+z''^{2})^{p}+(2\pi
l_{s}^{2})^{2}\sum_{n}^{p}\frac{n(n-1)}{n!}\left(-\frac{F}{\beta^{2}}\right)^{n}}F_{2}\nonumber\\&&F_{2}=
F_{1}\sqrt{1+\frac{k^{2}}{F_{1}^{2}\sigma^{2p}}} \label{m27}
\end{eqnarray}
and by repeating these calculations for $p$ times, we obtain:

\begin{eqnarray}
H_{p}&=& T_{Mp}\int d\sigma \sigma^{p}
\sqrt{(1+z'^{2}F(z)+z''^{2})^{p}+(2\pi
l_{s}^{2})^{2}\sum_{n=0}^{p}\frac{n(n-1)....(n-n)}{n!}\left(-\frac{F}{\beta^{2}}\right)^{n}}F_{p}\nonumber\\&=&
T_{Mp}\int d\sigma \sigma^{p}
(\sqrt{1+z'^{2}F(z)+z''^{2}})^{p}F_{tot}\nonumber\\
F_{tot}&=&\sqrt{1+\frac{k^{2}}{F_{p-1}^{2}\sigma^{2p}}}\sqrt{1+\frac{k^{2}}{F_{p-2}^{2}\sigma^{2p}}}...
\sqrt{1+\frac{k^{2}}{F_{1}^{2}\sigma^{2p}}}\sqrt{1+\frac{k^{2}}{\sigma^{2p}}}\nonumber\\F_{n}&=&F_{n-1}
\sqrt{1+\frac{k^{2}}{F_{n-1}^{2}\sigma^{2p}}} \label{m28}
\end{eqnarray}

By growing branes ($\sigma\rightarrow \infty$), the canonical
density ($\Pi$)in equation (\ref{m26}) becomes small. This is
because that this momentum relates to effect of one graviton on
total size of one brane and consequently, by increasing the length
of one brane, this effect decreases. However, by joining M1-branes
to each other, the number of gravitons on the branes increases and
total momentum density of gravity which is the sum over the
momentum densities of gravitons enhances. Thus, this total
momentum becomes large and plays the main role in evolution of
universe branes. At this stage, for the case of
$\frac{k}{\sigma^{p}}\ll1$, we can reproduce the Hamiltonian of
$Mp$-brane by multiplying the Hamiltonians of $M1$'s:

\begin{eqnarray}
H_{p}&=& T_{Mp}\int d\sigma \sigma^{p}
(\sqrt{1+z'^{2}F(z)+z''^{2}})^{p}F_{tot}\nonumber\\
k&=&k'^{p}\rightarrow
F_{tot}=\sqrt{1+\frac{k^{2}}{F_{p-1}^{2}\sigma^{2p}}}\sqrt{1+\frac{k^{2}}{F_{p-2}^{2}\sigma^{2p}}}...
\sqrt{1+\frac{k^{2}}{F_{1}^{2}\sigma^{2p}}}\sqrt{1+\frac{k^{2}}{\sigma^{2p}}}\nonumber\\&\simeq&
\sigma^{p}\sqrt{1+\frac{k'^{2p}}{\sigma^{2p}}+p\frac{k'^{2p-1}}{\sigma^{2p-1}}+..}=\sqrt{\left(1+\frac{k'^{2}}{\sigma^{2}}\right)^{p}}
\Rightarrow \nonumber\\&& H_{p}= T_{Mp}\int d\sigma \sigma^{2p}
(\sqrt{1+z'^{2}F(z)+z''^{2}})^{p}(\sqrt{1+\frac{k'^{2}}{\sigma^{2}}})^{p}=T_{Mp}\int
d^{p}\sigma \sigma^{p}
(\sqrt{1+z'^{2}})^{p}\left(\sqrt{1+\frac{k'^{2}}{\sigma^{2}}}\right)^{p}\Rightarrow
\nonumber\\ H_{p}&=&\Bigl(T_{M1}\int d\sigma \sigma
\sqrt{1+z'^{2}F(z)+z''^{2}}\sqrt{1+\frac{k'^{2}}{\sigma^{2}}}\Bigr)^{p}
=H_{1}^{p}  \nonumber\\
H_{1}&=&T_{M1}\int d\sigma \sigma
\sqrt{1+z'^{2}F(z)+z''^{2}}\sqrt{1+\frac{k'^{2}}{\sigma^{2}}}\label{m29}
\end{eqnarray}
where we have used of this assumption that $T_{Mp}=(T_{M1})^{p}$.
As can be seen from the above equation, each $Mp$-brane can be
constructed of $p M1$-brane. Also, we can show that each
$M1$-brane produces a wormhole. For this end, using the above
Hamiltonian and assuming that the acceleration of branes be
smaller than the velocity of branes in transverse dimension
($z''\ll z'$) and $F(z)\sim z^{2}$, we derive the following
equation of motion $z$ for any M1:

\begin{eqnarray}
-z_{M1}'&=&(\frac{V_{1}(\sigma)^{2}}{V_{1}(\sigma_{0})^{2}}-1)^{-1/2}z
\nonumber\\ V_{1}&=&\sigma F_{1}=\sigma
\sqrt{1+\frac{k'^{2}}{\sigma^{2}}}\label{m30}
\end{eqnarray}
The solution of this equation is:

\begin{eqnarray}
&&z_{M1}= e^{\int_{\sigma}^{\infty}
d\sigma'(\frac{V_{1}(\sigma')^{2}}{V_{1}(\sigma_{0})^{2}}-1)^{-1/2}
} \label{m31}
\end{eqnarray}
Thus, the separation distance between two branes is:
\begin{eqnarray}
&&\Delta_{M1}=2z_{M1}= 2 e^{\int_{\sigma}^{\infty}
d\sigma'(\frac{V_{1}(\sigma')^{2}}{V_{1}(\sigma_{0})^{2}}-1)^{-1/2}
} \label{m32}
\end{eqnarray}
where $\sigma_{0}$ is the throat of wormhole between two
$M1$-branes of two different branes. Thus, each $Mp$-brane is
constructed from $p M1$-branes which each of them produces a
wormhole and connects with the M1 of other branes.

Now, we can derive the relevant action for $Mp$-branes by
multiplying the action of p M1-branes by using the antisymmetric
form $\delta$:

\begin{eqnarray}
&& S_{Mp} = -T_{Mp} \int dt \L_{Mp}  \nonumber\\&& \L_{Mp}=\det(M)
\quad L_{M1,i}=L^{b_{i}}_{a_{i}}=\det(M_{i})\sim M_{i}~~~
\mbox{where,}
~~~~\det(M)=\delta^{a_{1},a_{2}...a_{n}}_{b_{1}b_{2}....b_{n}}M^{b_{1}}_{a_{1}}...M^{b_{p}}_{a_{p}}\Rightarrow
\nonumber\\&&
\L_{Mp}=\det(M)=\delta^{a_{1},a_{2}...a_{n}}_{b_{1}b_{2}....b_{n}}L^{b_{1}}_{a_{1}}...L^{b_{p}}_{a_{p}}
\nonumber~~~ \mbox{where,}
~~~~\delta^{a_{1},a_{2}...a_{n}}_{b_{1}b_{2}....b_{n}}
\delta^{\rho_{1}\sigma_{1}}_{\mu_{1}\nu_{1}}...\delta^{\rho_{p}\sigma_{p}}_{\mu_{p}\nu_{p}}
=\delta^{\rho_{1}\sigma_{1}...\rho_{p}\sigma_{p}}_{\mu_{1}\nu_{1}...\mu_{p}\nu_{p}}\nonumber\\&&
\sqrt{-g}=\sqrt{-\det(g)}=\sqrt{-\det(g_{1}g_{2}...g_{p})}=\sqrt{-\det(g_{1})\det(g_{2})...\det(g_{p})}\nonumber\\
&&\nonumber\\&&\nonumber\\&&
\end{eqnarray}
so we have
\begin{eqnarray}
S_{Mp} &=& -(T_{M1})^{p} \int dt
\delta^{a_{1},a_{2}...a_{n}}_{b_{1}b_{2}....b_{n}}L^{b_{1}}_{a_{1}}...L^{b_{p}}_{a_{p}}=\nonumber\\&&
-(T_{M1})^{p} \int dt \int d^{p}\sigma
\delta^{a_{1},a_{2}...a_{n}}_{b_{1}b_{2}....b_{n}}\nonumber\\
&&
\Bigl[\sqrt{-g_{1}}\Big(\delta^{\rho_{1}\sigma_{1}}_{\mu_{1}\nu_{1}}[\kappa_{\mu_{1}}^{\rho_{1}}\kappa_{\nu_{1}}^{\sigma_{1}}
\sum (X^{i})^{2}+
(\partial_{\lambda}\kappa_{\mu_{1}}^{\rho_{1}}\partial^{\lambda}\kappa_{\nu_{1}}^{\sigma_{1}})]-
m_{g}^{2}\delta^{\rho_{1}\sigma_{1}}_{\mu_{1}\nu_{1}}(R_{\rho_{1}\sigma_{1}}^{\mu_{1}\nu_{1}}+
[\kappa_{\mu_{1}}^{\rho_{1}}\kappa_{\nu_{1}}^{\sigma_{1}}\sum
(X^{i})^{2}+
(\partial_{\lambda}\kappa_{\mu_{1}}^{\rho_{1}}\partial^{\lambda_{1}}\kappa_{\nu_{1}}^{\sigma_{1}})])\Big)\Bigr]^{b_{1}}_{a_{1}}\times
\nonumber\\&&.......\times\nonumber\\
&&
\Biggl[\sqrt{-g_{p}}\Big(\delta^{\rho_{p}\sigma_{p}}_{\mu_{p}\nu_{p}}[\kappa_{\mu_{p}}^{\rho_{p}}\kappa_{\nu_{p}}^{\sigma_{p}}
\sum (X^{i})^{2}+
(\partial_{\lambda}\kappa_{\mu_{p}}^{\rho_{p}}\partial^{\lambda}\kappa_{\nu_{p}}^{\sigma_{p}})]-
m_{g}^{2}\delta^{\rho_{p}\sigma_{p}}_{\mu_{p}\nu_{p}}(R_{\rho_{p}\sigma_{p}}^{\mu_{p}\nu_{p}}+[\kappa_{\mu_{p}}^{\rho_{p}}
\kappa_{\nu_{p}}^{\sigma_{p}} \sum (X^{i})^{2}+
(\partial_{\lambda}\kappa_{\mu_{p}}^{\rho_{p}}\partial^{\lambda_{p}}
\kappa_{\nu_{p}}^{\sigma_{p}})])\Big)\Biggr]^{b_{p}}_{a_{p}}\nonumber\\&=&-(T_{Mp})\int
dt \int d^{p}\sigma
\Biggl[\sqrt{-g}\Big(\sum_{n=1}^{p}\delta^{\rho_{1}\sigma_{1}...\rho_{n}\sigma_{n}}_{\mu_{1}\nu_{1}...\mu_{n}\nu_{n}}F(X)^{p}
\kappa^{\mu_{1}}_{\rho_{1}}\kappa^{\nu_{1}}_{\sigma_{1}}
..\kappa^{\mu_{n}}_{\rho_{n}}\kappa^{\nu_{n}}_{\sigma_{n}}-
\sum_{n=1}^{p}m_{g}^{2n}\delta^{\rho_{1}\sigma_{1}...\rho_{n}\sigma_{n}}_{\mu_{1}\nu_{1}...\mu_{n}\nu_{n}}
R_{\rho_{1}\sigma_{1}}^{\mu_{1}\nu_{1}}...R_{\rho_{n}\sigma_{n}}^{\mu_{n}\nu_{n}}
\nonumber\\&&-
\sum_{n=1}^{p}m_{g}^{2n}F(X)^{p}\delta^{\rho_{1}\sigma_{1}...\rho_{n}\sigma_{n}}_{\mu_{1}\nu_{1}...\mu_{n}\nu_{n}}
\kappa^{\mu_{1}}_{\rho_{1}}\kappa^{\nu_{1}}_{\sigma_{1}}....\kappa^{\mu_{n}}_{\rho_{n}}\kappa^{\nu_{n}}_{\sigma_{n}}+
\sum_{n=1}^{p}(1-m_{g}^{2})\delta^{\rho_{1}\sigma_{1}...\rho_{n}\sigma_{n}}_{\mu_{1}\nu_{1}...\mu_{n}\nu_{n}}
\partial_{\lambda}\kappa^{\mu_{1}}_{\rho_{1}}\partial^{\lambda}\kappa^{\nu_{1}}_{\sigma_{1}}
..\partial_{\lambda}\kappa^{\mu_{n}}_{\rho_{n}}\partial^{\lambda}\kappa^{\nu_{n}}_{\sigma_{n}}+...
\Big)\Biggr] \label{m33}
\end{eqnarray}
This action includes all terms in nonlinear gravity theories like
Lovelock \cite{q32,q33} and massive gravity \cite{q24,q25,q26}. In
addition,  some extra terms are predicted only in this
model.  Now, we calculate
the number of degrees freedom on the universe brane and in a bulk.
Previously, we show that two form gauge fields are the main cause
of appearance of wormhole between $M$-branes. Thus, difference
between number of degrees freedom on the brane and bulk is due to
these fields. Using equations (\ref{m9}) and (\ref{m15}) and
assuming $A^{22}\sim g^{22}\sim l_{1}$ and $A^{ii}=g_{ij}=0$, we
get:

\begin{eqnarray}
&& A^{00}=g^{00}=-1 \quad A^{22}\sim g^{22}\sim l_{1} \quad
A^{ij}=0,i,j\neq0,2 \nonumber\\&& N_{sur}-N_{bulk}=2(T_{Mp}) \int
dt \int d^{p}\sigma
\Biggl[\sqrt{-g}\left(\sum_{n=1}^{p}\delta^{\rho_{1}\sigma_{1}...\rho_{n}\sigma_{n}}_{\mu_{1}\nu_{1}...\mu_{n}\nu_{n}}
\langle
F^{\rho_{1}\sigma_{1}}\smallskip_{\lambda},F^{\lambda}\smallskip_{\mu_{1}\nu_{1}}\rangle...\langle
F^{\rho_{n}\sigma_{n}}\smallskip_{\lambda},F^{\lambda}\smallskip_{\mu_{n}\nu_{n}}\rangle \right)\Biggr]=\nonumber\\
&&2(T_{Mp}) \int dt \int d^{p}\sigma
\Biggl[\sqrt{-g}\left(\sum_{n=1}^{p}\delta^{\rho_{1}\sigma_{1}...\rho_{n}\sigma_{n}}_{\mu_{1}\nu_{1}...\mu_{n}\nu_{n}}
R_{\rho_{1}\sigma_{1}}^{\mu_{1}\nu_{1}}...R_{\rho_{n}\sigma_{n}}^{\mu_{n}\nu_{n}}\right)\Biggr]\simeq
\nonumber\\
&&(T_{Mp})V
\sum_{n}\Biggl[\frac{l_{0}^{n}}{2^{n}(2n-1)(t_{s}-t)^{2n-1}}\Bigl[1-\frac{t_{s}^{1/2}}{\sqrt{t_{s}-t}}\Bigr]^{2n}-
\frac{l_{d,0}^{2n}}{t_{s}^{2n}}(t_{s}-t)^{n}\Biggr]e^{\frac{-2nt}{t_{s}}}
\label{m34}
\end{eqnarray}
where $V$ is the volume of brane, $p=3$ is related to our universe
and the number $2$ is related to exchanging graviton between two
branes which produce two sections of a wormhole. We also have used
of this fact that
$m_{g}^{2}=[(\lambda)^{2}\det([X^{j}_{\alpha}T^{\alpha},X^{k}_{\beta}T^{\beta},X^{k'}_{\gamma}T^{\gamma}])]
\sim 1+l_{d}^{3}$. This equation shows that by approaching branes,
difference between number of degrees of freedom increases and this
causes to the growth of branes and universe expansion. We also
have:

\begin{eqnarray}
&& A^{00}=g^{00}=-1 ~ ~  A^{22}\sim g^{22}\sim l_{1} ~ ~ A^{ij}=0
\smallskip \mbox{ i,j$\neq$0,2} \quad X^{2}=l_{1} ~ ~ X^{i}=0
\smallskip \mbox{i$\neq$ 2}
\nonumber\\&&N_{sur}+N_{bulk}=E_{Mp}+E_{anti-Mp}=\nonumber\\&&-2(T_{Mp})
\int dt \int d^{p}\sigma
\Biggl[\sqrt{-g}\Bigl(\sum_{n=1}^{p}\delta^{\rho_{1}\sigma_{1}...\rho_{n}\sigma_{n}}_{\mu_{1}\nu_{1}...
\mu_{n}\nu_{n}}F(X)^{p}
\kappa^{\mu_{1}}_{\rho_{1}}\kappa^{\nu_{1}}_{\sigma_{1}}
..\kappa^{\mu_{n}}_{\rho_{n}}\kappa^{\nu_{n}}_{\sigma_{n}}-\nonumber\\&&
\sum_{n=1}^{p}m_{g}^{2n}\delta^{\rho_{1}\sigma_{1}...\rho_{n}\sigma_{n}}_{\mu_{1}\nu_{1}...\mu_{n}\nu_{n}}
R_{\rho_{1}\sigma_{1}}^{\mu_{1}\nu_{1}}...R_{\rho_{n}\sigma_{n}}^{\mu_{n}\nu_{n}}
-\nonumber\\&&
\sum_{n=1}^{p}m_{g}^{2n}F(X)^{p}\delta^{\rho_{1}\sigma_{1}...\rho_{n}\sigma_{n}}_{\mu_{1}\nu_{1}...\mu_{n}\nu_{n}}
\kappa^{\mu_{1}}_{\rho_{1}}\kappa^{\nu_{1}}_{\sigma_{1}}....\kappa^{\mu_{n}}_{\rho_{n}}\kappa^{\nu_{n}}_{\sigma_{n}}+
\nonumber\\&&\sum_{n=1}^{p}(1-m_{g}^{2n})\delta^{\rho_{1}\sigma_{1}...\rho_{n}\sigma_{n}}_{\mu_{1}\nu_{1}...\mu_{n}\nu_{n}}
\partial_{\lambda}\kappa^{\mu_{1}}_{\rho_{1}}\partial^{\lambda}\kappa^{\nu_{1}}_{\sigma_{1}}
..\partial_{\lambda}\kappa^{\mu_{n}}_{\rho_{n}}\partial^{\lambda}\kappa^{\nu_{n}}_{\sigma_{n}}+...\Bigr)\Biggr]
= \nonumber\\&& (T_{Mp})V\sum_{n}
\Biggl[\frac{6^{n}l_{1,+}^{2n}}{(t_{s}-t)^{8n}}+\frac{l_{1,+}^{2n}l_{0}^{2n}}{(t_{s}-t)^{2n}}+
\frac{2^{n}l_{1,+}^{2n}}{(t_{s}-t)^{6n}}\Biggr]\Biggl[1-\frac{t_{s}}{(t_{s}-t)}\Biggr]^{2n}e^{-2nl_{0}(t_{s}-t)}
+ \nonumber\\&&(T_{Mp})V
\sum_{n}\Bigl[\frac{l_{d,0}^{2n}}{t_{s}^{2n}}(t_{s}-t)^{n}\Bigr]e^{\frac{-2nt}{t_{s}}}\label{m35}
\end{eqnarray}
Solving equations (\ref{m34})and (\ref{m35}), we can obtain the
explicit form of  degrees of freedom of bulk and brane:

\begin{eqnarray}
N_{sur}=&&(T_{Mp})V\sum_{n=1}^{p}\Biggl[\frac{6^{n}l_{1,+}^{2n}}{(t_{s}-t)^{8n}}+\frac{l_{1,+}^{2n}l_{0}^{2n}}{(t_{s}-t)^{2n}}+
\frac{2^{n}l_{1,+}^{2n}}{(t_{s}-t)^{6n}}\Biggr]\Biggl[1-\frac{t_{s}}{(t_{s}-t)}\Biggr]^{2n}e^{-2nl_{0}(t_{s}-t)}
\nonumber\\&&+ (T_{Mp})V
\sum_{n=1}^{p}\Biggl[\frac{l_{0}^{n}}{2^{n}(2n-1)(t_{s}-t)^{2n-1}}\Biggl[1-\frac{t_{s}^{1/2}}{\sqrt{t_{s}-t}}\Biggr]^{2n}\Biggr]
e^{\frac{-2nt}{t_{s}}} \label{m36}
\end{eqnarray}

\begin{eqnarray}
N_{bulk}=2(T_{Mp})V
\sum_{n=1}^{p}\Biggl[\frac{l_{d,0}^{2n}}{t_{s}^{2n}}(t_{s}-t)^{n}\Biggr]e^{\frac{-2nt}{t_{s}}}\label{m37}
\end{eqnarray}
Clearly, at the beginning ($t=0$),  the number of degrees of
freedom on the surface of brane is zero; while, by evolving the
time and approaching branes towards each other, the number of
degrees of freedom on the brane increases and tends to infinity
(see Figure 1. Left). On the other hand, the number of degrees of
freedom in the bulk decreases with time and shrinks to zero at
colliding point ($t=t_{s}$)(see figure.1 Right ) .

\begin{figure*}[thbp]
\begin{tabular}{rl}
\includegraphics[width=6.0cm]{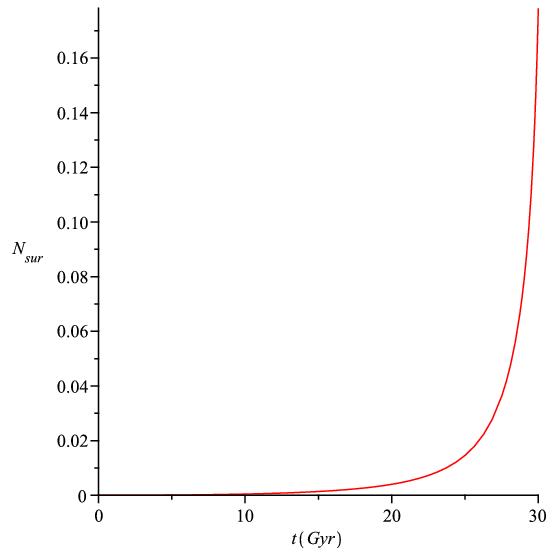}&
\includegraphics[width=6.0cm]{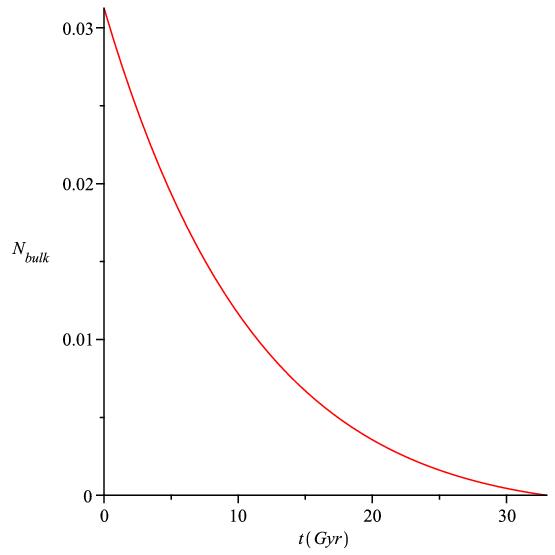}\\
\end{tabular}
\caption{ (Left) $N_{sur}$  is  increasing  from zero at $t=0$ to
infinity at $t=t_s$ . ( Right)  $N_{bulk}$ is decreasing from
certain  value $ at t=0$  to zero $t= t_s$. We have assumed $p=3$
for 3+1 dimensional M3 which our universe is located on it and
time of collision between branes $t_s=33 Gyr $. }
\end{figure*}


\section{Contraction branch of cosmic space in Padmanabhan model}\label{o2}
Until now, we have shown that by approaching branes, their size
grows causing  the expansion of the universe. Now, we will show
that near the collision point, branes compact, universe contracts
and gravity changes to anti-gravity. This causes that branes get
away from each other. To this end, let us to consider equations
(\ref{m13}) and(\ref{m15}) near the colliding point:

\begin{eqnarray}
&& t\rightarrow t_{s}\Rightarrow l_{d}\sim
\frac{l_{d,0}}{t_{s}^{3/2}}(t_{s}-t)^{3/2}\rightarrow
0\Rightarrow\nonumber\\&& l_{1}\sim
\frac{l_{1,+}}{(t_{s}-t)^{2}}\left(1+\frac{1}{l_{1,0}(t_{s}-t)}\right)e^{-l_{0}(t_{s}-t)}
\rightarrow l'_{1}l_{d}\sim \frac{1}{(t_{s}-t)^{3}}\rightarrow
\infty \Rightarrow \nonumber\\&&
 D_{l_{d},l_{1}}=[l_{1}^{2}+(l'_{d})^{2}+(l''_{d})^{2}(1+l_{d}^{2})^{-1}]\left[1-\frac{1}{l_{d}^{3}}\right]-
 (l'_{1})^{2}(l_{d})^{2}\ll 0\label{m38}
\end{eqnarray}

This equation shows that by closing M1-branes to each other,
$D_{l_{d},l_{1}}\ll 0$ and thus the expression under $\sqrt{}$ in
the action in equation (\ref{m13}) becomes negative. This means
that the square energy of system becomes negative and some
tachyonic states are produced. To solve this problem, M1-branes
compact and the sign of gravity changes. To show this, we use of
the method in \cite{q12,q23} and define ${\displaystyle
<X^{10}>=\frac{R}{l_{p}^{3/2}}}$ where $l_{p}$ is the Planck
length. We can write:

\begin{eqnarray}
&& [X^{a},X^{b},X^{c}]=F^{abc} \quad [X^{a},X^{b}]=F^{ab}
 \nonumber \\
&&F_{abc}=\partial_{a} A_{bc}-\partial_{b} A_{ca}+\partial_{c}
A_{ab} \quad F_{ab}=\partial_{a} A_{b}-\partial_{b} A_{a}\nonumber \\ \nonumber \\
&& \nonumber \\
\Sigma_{a,b,c=0}^{10} \langle
F^{abc},F_{abc}\rangle&=&\Sigma_{a,b,c=0}^{10}
\langle[X^{a},X^{b},X^{c}],[X_{a},X_{b},X_{c}]\rangle\nonumber \\
&=&
- \Sigma_{a,b,c,a'b'c'=0}^{10}\varepsilon_{abcD}\varepsilon_{a'b'c'G}^{D}X^{a}X^{b}X^{c}X_{a'}X_{b'}X_{c'} \nonumber \\
&=& -
6\Sigma_{a,b,a',b'=0}^{9}\varepsilon_{ab10D}\varepsilon_{a'b'10}^{D}X^{a}X^{b}X^{10}X_{a'}X_{b'}X_{10}
\nonumber \\
&-&  6\Sigma_{a,b,c,a',b',c'=0,\neq
10}^{9}\varepsilon_{abcD}\varepsilon_{a'b'c'}^{D}X^{a}X^{b}X^{c}X_{a'}X_{b'}X_{c'}
\nonumber \\&=&
 - 6\left(\frac{R^{2}}{l_{p}^{3}}\right)\Sigma_{a,b,a',b'=0}^{9}\varepsilon_{ab10D}\varepsilon_{a'b'10}^{D}X^{a}X^{b}X_{a'}X_{b'} \nonumber \\
&-&
6\Sigma_{a,b,c,a',b',c'=0,\neq10}^{9}\varepsilon_{abcD}\varepsilon_{a'b'c'}^{D}X^{a}X^{b}X^{c}X_{a'}X_{b'}X_{c'}  \nonumber \\
&=& - 6\left(\frac{R^{2}}{l_{p}^{3}}\right)\Sigma_{a,b=0}^{9}[X^{a},X^{b}]^{2} \nonumber \\
&-&  6\Sigma_{a,b,c,a',b',c'=0,\neq10}^{9}\varepsilon_{abcD}\varepsilon_{a'b'c'}^{D}X^{a}X^{b}X^{c}X_{a'}X_{b'}X_{c'} \nonumber \\
&=& -
6\left(\frac{R^{2}}{l_{p}^{3}}\right)\Sigma_{a,b=0}^{9}F^{ab}F_{ab}
+ E_{Extra} \label{m39}
\end{eqnarray}
This equation shows that two form fields in eleven dimensional
space-time transit to one form field as due to compactification
and the sign of self energy changes. Using equation (\ref{m39}),
we can replace all two-form terms in gravity theories by one-form
terms:

\begin{eqnarray}
&&A_{b}=e_{b}\quad F_{ab}=\partial_{a} e_{b}-\partial_{b} e_{a}\quad \kappa^{a}\smallskip_{b}=\partial^{a} e_{b}\nonumber \\
&&\Sigma_{\rho,\sigma,\mu,\nu=0}^{10}R^{\rho\sigma}_{\mu\nu}=\Sigma_{\rho,\sigma,\mu,\nu,\lambda=0}^{10}\langle
F^{\rho\sigma}\smallskip_{\lambda},F^{\lambda}\smallskip_{\mu\nu}\rangle=\Sigma_{\rho,\sigma,\mu,\nu,\lambda=0}^{10}
\langle[X^{\rho},X^{\sigma},X_{\lambda}],[X^{\lambda},X_{\mu},X_{\nu}]\rangle\Rightarrow
\nonumber\\&&
-6\left(\frac{R^{2}}{l_{p}^{3}}\right)\Sigma_{\rho,\sigma,\mu,\nu=0}^{9}[X^{\rho},X^{\sigma}][X_{\mu},X_{\nu}]=
-6\left(\frac{R^{2}}{l_{p}^{3}}\right)
\Sigma_{\rho,\sigma,\mu,\nu=0}^{9}F^{\rho\sigma}F_{\mu,\nu}=\nonumber\\&&-6(\frac{R^{2}}{l_{p}^{3}})
\Sigma_{\rho,\sigma,\mu,\nu=0}^{9}\delta^{\rho\sigma}_{\rho'\sigma'}\delta_{\mu\nu}^{\mu'\nu'}\partial^{\rho'}
e^{\sigma'}\partial_{\mu'}
e_{\nu'}=-6\left(\frac{R^{2}}{l_{p}^{3}}\right)
\Sigma_{\rho,\sigma,\mu,\nu=0}^{9}\delta^{\rho\sigma}_{\rho'\sigma'}\delta_{\mu\nu}^{\mu'\nu'}\kappa^{\rho'\sigma'}
\kappa_{\mu'\nu'}\nonumber\\&&
\Sigma_{\rho,\sigma,\mu,\nu=0}^{10}R=-6\left(\frac{R^{2}}{l_{p}^{3}}\right)
\Sigma_{\rho,\sigma,\mu,\nu=0}^{9}\delta^{\mu\nu}_{\rho'\sigma'}\delta_{\mu\nu}^{\mu'\nu'}\kappa^{\rho'\sigma'}
\kappa_{\mu'\nu'}\label{m40}
\end{eqnarray}
With the help of these relations, we can show that the sign of
Lovelock gravity changes:
\begin{eqnarray}
\sum_{n=1}^{p}m_{g}^{2n}\delta^{\rho_{1}\sigma_{1}...\rho_{n}\sigma_{n}}_{\mu_{1}\nu_{1}...\mu_{n}\nu_{n}}
R_{\rho_{1}\sigma_{1}}^{\mu_{1}\nu_{1}}...R_{\rho_{n}\sigma_{n}}^{\mu_{n}\nu_{n}}\Rightarrow
-\sum_{n=1}^{p}m_{g}^{2n}\left(\frac{R^{2n}}{l_{p}^{3n}}\right)
\delta^{\rho_{1}\sigma_{1}...\rho_{n}\sigma_{n}}_{\mu_{1}\nu_{1}...\mu_{n}\nu_{n}}\kappa^{\mu_{1}}_{\rho_{1}}
\kappa^{\nu_{1}}_{\sigma_{1}}
..\kappa^{\mu_{n}}_{\rho_{n}}\kappa^{\nu_{n}}_{\sigma_{n}}\label{m41}
\end{eqnarray}
This equation shows that by compacting Mp-brane, nonlinear
theories like Lovelock gravity converts to other type of non-linear
gravity theories with opposite sign. This means that by compacting
branes, gravity changes to anti-gravity.  We can study other
effects of compactifications of Mp-brane by extending the
relations in (\ref{m41}):

\begin{eqnarray}
&&\partial_{a}X^{i}\partial_{a}X^{i}\sum
(X^{j})^{2}=\Sigma_{a,i,j=0}^{10}\langle[X^{a},X^{i},X^{j}],[X_{a},X_{i},X_{j}]\rangle\Rightarrow
\nonumber\\&&
-6\left(\frac{R^{2}}{l_{p}^{3}}\right)\Sigma_{i,j=0}^{9}[X^{a},X^{j}][X_{a},X_{j}]=-6\left(\frac{R^{2}}{l_{p}^{3}}\right)
\partial_{a}X^{i}\partial_{a}X^{i}\Rightarrow
F(X)=\sum
(X^{j})^{2}=1\nonumber\\
&&\partial_{a}\partial_{b}
X^{i}\partial_{a}\partial_{b}X^{i}=\Sigma_{a,b,i=0}^{10}\langle[X^{a},X^{b},X^{i}],[X_{a},X_{b},X_{i}]\rangle\Rightarrow
\nonumber\\&&
-6\left(\frac{R^{2}}{l_{p}^{3}}\right)\Sigma_{i,b=0}^{9}[X^{b},X^{i}][X_{b},X_{i}]=-6\left(\frac{R^{2}}{l_{p}^{3}}\right)\partial_{a}
X^{i}\partial_{a}X^{i}\label{m42}
\end{eqnarray}
When scalars attached to branes give the index of branes, they play
the role of graviton. In these conditions, using equation
(\ref{m42}), we can obtain the following relations.

\begin{eqnarray}
&&X^{i}\rightarrow e^{c}\Rightarrow \partial_{a}\partial_{b}
X^{i} \rightarrow \partial_{a} \kappa^{c}_{b} \nonumber\\
&&\partial_{a} \kappa^{c}_{b}\partial_{a} \kappa^{c}_{b}
\rightarrow\partial_{a}\partial_{b}
X^{i}\partial_{a}\partial_{b}X^{i}=\Sigma_{a,b,i=0}^{10}\langle[X^{a},X^{b},X^{i}],[X_{a},X_{b},X_{i}]\rangle\Rightarrow
\nonumber\\&&
-6\left(\frac{R^{2}}{l_{p}^{3}}\right)\Sigma_{i,b=0}^{9}[X^{b},X^{i}][X_{b},X_{i}]=-6\left(\frac{R^{2}}{l_{p}^{3}}\right)\partial_{a}
X^{i}\partial_{a}X^{i}\rightarrow-6\left(\frac{R^{2}}{l_{p}^{3}}\right)\kappa^{c}_{b}\kappa^{c}_{b}\nonumber\\
&&\nonumber\\
&&\Rightarrow
\sum_{n=1}^{p}\delta^{\rho_{1}\sigma_{1}...\rho_{n}\sigma_{n}}_{\mu_{1}\nu_{1}...\mu_{n}\nu_{n}}
\partial_{\lambda}\kappa^{\mu_{1}}_{\rho_{1}}\partial^{\lambda}\kappa^{\nu_{1}}_{\sigma_{1}}
..\partial_{\lambda}\kappa^{\mu_{n}}_{\rho_{n}}\partial^{\lambda}\kappa^{\nu_{n}}_{\sigma_{n}}\nonumber\\
&&\rightarrow
-6\left(\frac{R^{2}}{l_{p}^{3}}\right)\Sigma_{i,b=0}^{9}\delta^{\rho_{1}\sigma_{1}...\rho_{n}\sigma_{n}}_{\mu_{1}
\nu_{1}...\mu_{n}\nu_{n}}\kappa^{\mu_{1}}_{\rho_{1}}\kappa^{\nu_{1}}_{\sigma_{1}}
..\kappa^{\mu_{n}}_{\rho_{n}}\kappa^{\nu_{n}}_{\sigma_{n}}\label{m43}
\end{eqnarray}
Substituting equations (\ref{m41}) and (\ref{m43}) into action
(\ref{m35}), we observe that four non-linear terms of five will be
removed by each other and only one term remains:

\begin{eqnarray}
&&S_{Mp} = -(T_{Mp}) \int dt \int d^{p}\sigma
\Biggl[\sqrt{-g}\Big(\sum_{n=1}^{p}m_{g}^{2n}\Bigl(1+6^{n}\Bigl(\frac{R^{2n}}{l_{p}^{3n}}\Bigr)\Bigr)\delta^{\rho_{1}\sigma_{1}...\rho_{n}
\sigma_{n}}_{\mu_{1}\nu_{1}...\mu_{n}\nu_{n}}\kappa^{\mu_{1}}_{\rho_{1}}\kappa^{\nu_{1}}_{\sigma_{1}}
..\kappa^{\mu_{n}}_{\rho_{n}}\kappa^{\nu_{n}}_{\sigma_{n}}\Big)\Biggr]
\label{m44}
\end{eqnarray}
where $m_{g}^{2}=[(\lambda)^{2}\det([X^{j},X^{k}])] $ is the
square of graviton mass. This equation show that compacting
Mp-branes gives rise to  anti-gravity. For example, it is

\begin{eqnarray}
&&\sqrt{-g}R\rightarrow -m_{g}^{2}\sqrt{-g}R\label{m45}
\end{eqnarray}
for general relativity. In fact, by approaching branes, they
compact, universe contracts and gravity changes to anti gravity.
In these conditions, branes are getting away from each other and
contraction branch ends. To show this, similar to the previous
section, we consider the action of M1-branes and then extend it to
higher dimensional compact branes. We can rewrite the action of
compacted M1-brane as
\cite{q12,q15,q16,q17,q18,q19,qq19,q20,q21,q22,q23}:

\begin{eqnarray}
S = - T_{M1}\int  d^{2}\sigma ~ STr \sqrt{-det(P_{ab}[ E_{mn}
+E_{mi}(Q^{-1}-\delta)^{ij}E_{jn}]+\lambda
F_{ab})det(Q^{i}_{j})}~~ \label{m46}
\end{eqnarray}
where
\begin{eqnarray}
E_{mn} = G_{mn} + B_{mn}, \qquad  Q^{i}_{j} = \delta^{i}_{j}
    + i\lambda[X^{j},X^{k}]E_{kj}, \qquad F_{ab}=\partial_{a} A_{b}-\partial_{b} A_{a} \label{m47}
\end{eqnarray}
where $\lambda=2\pi l_{s}^{2}$,
$G_{mn}=g_{mn}+\partial_{m}X^{i}\partial_{n'}X^{i} $ and $X^{i}$
are scalar fields with mass dimension. Using the above action and
and assuming that, as in  previous section, the separation distance
between two $M1$ be $l_{d}$ and the length of each $M1$ be
$l_{1}$, we can derive the relevant action for the interaction of
an $M1$ with an anti-$M1$-brane:

\begin{eqnarray}
&&  A^{a}\rightarrow l_{1} \quad X^{2}\rightarrow l_{d} \quad X^{0}=t \quad X^{i}=0, i\neq 0,2\nonumber\\
&& S=-T_{M1}\int d^{2}\sigma \sqrt{l_{d}^{2}+1}
\sqrt{[l_{1}^{2}+(l'_{d})^{2}]\Bigl[1-\frac{1}{l_{d}^{2}}\Bigr]+(l'_{1})^{2}}=\nonumber\\&&-T_{M1}\int
d^{2}\sigma V(l_{d}) \sqrt{D_{l_{d},l_{1}}}\nonumber\\&&
 V(l_{d})= \sqrt{l_{d}^{2}+1}\nonumber\\&&
 D_{l_{d},l_{1}}=[l_{1}^{2}+(l'_{d})^{2}]\Bigl[1-\frac{1}{l_{d}^{2}}\Bigr]+(l'_{1})^{2}\label{m48}
\end{eqnarray}
where the prime denotes the derivative respect to time. The
equations of motion extracted from the above action are:

\begin{eqnarray}
&&
\Biggl(\frac{(l'_{d})[1-\frac{1}{l_{d}^{2}}]}{\sqrt{D_{l_{d},l_{1}}}}\Biggr)'=\frac{1}{\sqrt{D_{l_{d},l_{1}}}}
\Biggl(2l_{d}^{-3}[(l'_{d})^{2}+l_{1}^{2}]l'_{d}
+\frac{V'}{V}[D_{l_{d},l_{1}}-(l'_{d})[1-\frac{1}{l_{d}^{2}}]]\Biggr)\nonumber\\&&
\Biggl(\frac{2(l'_{1})}{\sqrt{D_{l_{d},l_{1}}}}\Biggr)'=
\frac{1}{\sqrt{D_{l_{d},l_{1}}}}\Biggl(l_{1}\Biggl[1-\frac{1}{l_{d}^{2}}\Biggr]-\frac{V'}{V}[l'_{1}l'_{d}]\Biggr)\label{m49}
\end{eqnarray}
The approximate solutions of the above equations are:

\begin{eqnarray}
&& l_{1}\sim
\frac{l_{1,-}}{(t-t_{s})^{2}}\left(1+\frac{t_{s}}{(t-t_{s})}\right)e^{l_{0}(t_{s}-t)}
\nonumber\\&& l_{d}\sim
\frac{l_{d,-}}{t_{s}}(t-t_{s})\Biggl[1+2\ln\Biggl[1+\frac{(t-t_{s})}{t_{s}}\Biggr]\Biggr]\label{m50}
\end{eqnarray}
It is clear that at $t=t_{s}$, the separation of distance between
branes ($l_{d}=0$) is zero and the length of $M1$ is approximately
infinite; while, by passing time, the distance between $M1$-branes
increases and the length of branes decreases. We can examine the
correctness of these solutions near the colliding point that
branes are very closed to each other. In this case, the size of
branes before and after collision are approximately equal
($l_{1,before}(t=t_{s})=l_{1,after}(t=t_{s})$). Using this
assumption, equations in (\ref{m50}) reduce to following
equations:

\begin{eqnarray}
&&   l_{d}\sim 0 \quad l'_{d}\sim 0 \nonumber\\&&
\Rightarrow\frac{V'}{V}D_{l_{d},l_{1}}\sim
\frac{l_{d}l'_{d}}{l_{d}^{2}+1}\Bigl[[l_{1}^{2}+(l'_{d})^{2}]\Bigl[1-\frac{1}{l_{d}^{2}}\Bigr]+(l'_{1})^{2}\Bigr]\sim\nonumber\\&&
-2l_{d}^{-3}[(l'_{d})^{2}+l_{1}^{2}]l'_{d}+\frac{l'_{d}l_{1}^{2}}{l_{d}}\nonumber\\&&\text{And}\nonumber\\&&
l_{1,before}(t\rightarrow t_{s})=l_{1,after}(t\rightarrow
t_{s})\simeq |\frac{t_{s}}{(t-t_{s})^{3}}|\nonumber\\&&\Rightarrow
l_{1}^{2}+(l'_{1})^{2}=(l_{1})^{2}[1+\frac{(t-t_{s})^{2}}{t_{s}^{2}}]=(l_{1})^{2}\Biggl[[1+\frac{(t-t_{s})}{t_{s}}]^{2}-2\frac{(t-t_{s})}{t_{s}}\Biggr]\nonumber\\&&\Rightarrow
\Biggl(\frac{(l'_{d})[1-\frac{1}{l_{d}^{2}}]}{\sqrt{D_{l_{d},l_{1}}}}\Biggr)'=\frac{2}{1+\frac{(t-t_{s})}{t_{s}}}\nonumber\\&&\Rightarrow
 l_{d}\sim
\frac{l_{d,-}}{t_{s}}(t-t_{s})\Biggl[1+2\ln\Biggl[1+\frac{(t-t_{s})}{t_{s}}\Biggr]\Biggr]\nonumber\\&&\nonumber\\&&\nonumber\\&&\Biggl(\frac{2(l'_{1})}{\sqrt{D_{l_{d},l_{1}}}}\Biggr)'
\simeq \frac{2(l'_{1})}{\sqrt{D_{l_{d},l_{1}}}}\Rightarrow
\nonumber\\&&l_{1}\sim
\frac{l_{1,-}}{(t-t_{s})^{2}}\left(1+\frac{t_{s}}{(t-t_{s})}\right)e^{l_{0}(t_{s}-t)}
\label{1m50}
\end{eqnarray}

 These results can
be extended to higher dimensional branes. We have constructed the
action of (\ref{m44}) from compacting terms in action (\ref{m33}).
On the other hand, in previous section, we have proved that each
$Mp$-branes can be built from $p M1$-branes.

\begin{eqnarray}
&& S_{Mp} = -T_{Mp} \int dt
\delta^{a_{1},a_{2}...a_{n}}_{b_{1}b_{2}....b_{n}}L^{b_{1}}_{a_{1}}...L^{b_{p}}_{a_{p}}
\quad H\sim H_{1}^{p}\label{mm50}
\end{eqnarray}
This means that results of equation (\ref{m50}) can be generalized
to $Mp$-branes and we can choose the same lengths for all
dimensions of brane :

\begin{eqnarray}
&& l_{2}=...l_{p}=l_{1}\sim
\frac{l_{1,-}}{(t-t_{s})^{2}}\left(1+\frac{t_{s}}{(t-t_{s})}\right)e^{l_{0}(t_{s}-t)}
\label{mmm50}
\end{eqnarray}

At this stage, we can write the relations between the number of
degrees of freedom on the brane and in the bulk and the energy of
system. Until now, we have shown that one-form gauge fields
produce anti-gravity which are the main cause of inequality
between number of degrees of freedom on the brane and in a bulk.
Substituting equations (\ref{m41}), (\ref{m43}) and (\ref{m44}) in
equations (\ref{m34}) and (\ref{m35}), we obtain:

\begin{eqnarray}
&& N_{sur}-N_{bulk}=2(T_{Mp}) \int dt \int d^{p}\sigma
\Biggl[\sqrt{-g}\Bigl(\sum_{n=1}^{p}\delta^{\rho_{1}\sigma_{1}...\rho_{n}\sigma_{n}}_{\mu_{1}\nu_{1}...\mu_{n}\nu_{n}}
F^{\rho_{1}\sigma_{1}}F_{\mu_{1}\nu_{1}}...
F^{\rho_{n}\sigma_{n}}F_{\mu_{n}\nu_{n}}\Bigr)\Biggr]=\nonumber\\
&&2(T_{Mp})\int dt \int d^{p}\sigma
\Biggl[\sqrt{-g}\Bigl(\sum_{n=1}^{p}\delta^{\rho_{1}\sigma_{1}...\rho_{n}\sigma_{n}}_{\mu_{1}\nu_{1}...\mu_{n}\nu_{n}}
\kappa^{\mu_{1}}_{\rho_{1}}\kappa^{\nu_{1}}_{\sigma_{1}}
..\kappa^{\mu_{n}}_{\rho_{n}}\kappa^{\nu_{n}}_{\sigma_{n}}\Bigr)\Biggr]\label{m51}
\end{eqnarray}

\begin{eqnarray}
&&
N_{sur}+N_{bulk}=E_{compact-M3}+E_{compact-anti-M3}=\nonumber\\&&2(T_{Mp})
\int dt \int d^{p}\sigma
\Biggl[\sqrt{-g}\Bigl(\sum_{n=1}^{p}m_{g}^{2n}(1+6^{n}(\frac{R^{2n}}{l_{p}^{3n}}))\delta^{\rho_{1}\sigma_{1}...\rho_{n}
\sigma_{n}}_{\mu_{1}\nu_{1}...\mu_{n}\nu_{n}}\kappa^{\mu_{1}}_{\rho_{1}}\kappa^{\nu_{1}}_{\sigma_{1}}
..\kappa^{\mu_{n}}_{\rho_{n}}\kappa^{\nu_{n}}_{\sigma_{n}}\Bigr)\Biggr]\label{m52}
\end{eqnarray}
Solving the above equations, using equation (\ref{m50}) and
assuming $m_{g}^{2}=[(\lambda)^{2}\det([X^{j},X^{k}])]=1+l_{d}
^{2}$ and ${\displaystyle R=\frac{l_{p}^{\frac{3}{2}}}{6}}$ we obtain the surface
degrees of freedom and the one of bulk as follows:

\begin{eqnarray}
&&  A^{a}\rightarrow l_{1} \quad X^{2}\rightarrow l_{d} \quad
X^{0}=t \quad X^{i}=0, i\neq 0,2\nonumber\\&& N_{sur}=(T_{Mp})
\int dt \int d^{p}\sigma
\Biggl[\sqrt{-g}\Bigl(\sum_{n=1}^{p}(l_{d}^{2n}(1+6^{n}(\frac{R^{2n}}{l_{p}^{3n}}))+2)
\delta^{\rho_{1}\sigma_{1}...\rho_{n}\sigma_{n}}_{\mu_{1}\nu_{1}...\mu_{n}\nu_{n}}\kappa^{\mu_{1}}_{\rho_{1}}
\kappa^{\nu_{1}}_{\sigma_{1}}
..\kappa^{\mu_{n}}_{\rho_{n}}\kappa^{\nu_{n}}_{\sigma_{n}}\Bigr)\Biggr]\sim
\nonumber\\&&(T_{Mp})V\sum_{n=1}^{p}\Biggl[\frac{l_{d,-}^{2n}}{t_{s}^{2n}(2n-1)}\frac{1}{(t-t_{s})^{2n-1}}+
\frac{l_{d,-}^{2n}}{t_{s}^{2n}}t_{s}(t-t_{s})\ln
\Biggl[1+\frac{(t-t_{s})}{t_{s}}\Biggr]
\Biggl[1+2\ln\Biggl[1+\frac{(t-t_{s})}{t_{s}}\Biggr]\Biggr]^{2n-1}\Biggr]+
\nonumber\\&&(T_{Mp})V\sum_{n=1}^{p}\Biggl[\frac{l_{d,-}^{2n}}{t_{s}^{2n}(2n+1)}(t-t_{s})^{2n+1}\Biggr]\Biggl[1+(t-t_{s})
\ln\Biggl[1+\frac{(t-t_{s})}{t_{s}}\Biggr]+(t-t_{s})+\frac{t_{s}^{2}}{t-t_{s}}\Biggr]^{n}\label{m53}
\end{eqnarray}

\begin{eqnarray}
&& N_{bulk}=2(T_{Mp}) \int dt \int d^{p}\sigma
[\sqrt{-g}\Big(\sum_{n=1}^{p}(l_{d}^{2n})\delta^{\rho_{1}\sigma_{1}...\rho_{n}
\sigma_{n}}_{\mu_{1}\nu_{1}...\mu_{n}\nu_{n}}\kappa^{\mu_{1}}_{\rho_{1}}\kappa^{\nu_{1}}_{\sigma_{1}}
..\kappa^{\mu_{n}}_{\rho_{n}}\kappa^{\nu_{n}}_{\sigma_{n}}\Big)]\sim
\nonumber\\&&(T_{Mp})V\sum_{n=1}^{p}\Biggl[\frac{l_{d,-}^{2n}}{t_{s}^{2n}(2n+1)}(t-t_{s})^{2n+1}\Biggr]
\Biggl[1+(t-t_{s})\ln\Biggl[1+\frac{(t-t_{s})}{t_{s}}\Biggr]+(t-t_{s})+\frac{t_{s}^{2}}{t-t_{s}}\Biggr]^{n}\label{m54}
\end{eqnarray}
These equations show that at the colliding point
($t=t_{s}$), $m_{g}^{2}=1$, the number of degrees of freedom in the
bulk is zero ($N_{bulk}=0$) and the number of degrees of freedom
on the brane surface becomes infinite ($N_{sur}=\infty$). However
by passing time, degrees of freedom on the brane surface decrease
and shrinks to zero while, degrees of freedom in the bulk
increases (see figure 2).

\begin{figure*}[thbp]
\begin{tabular}{rl}
\includegraphics[width=6.0cm]{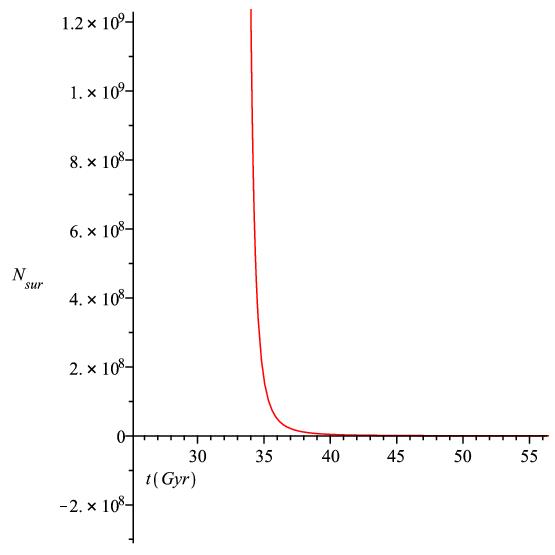}&
\includegraphics[width=6.0cm]{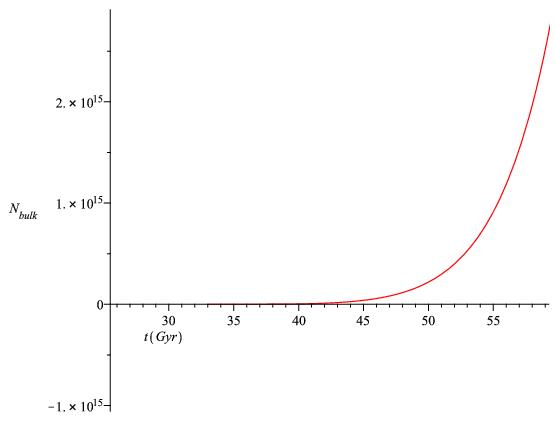}\\
\end{tabular}
\caption{ ( Left) $N_{sur}$  is  decreasing  from large value  at
$t=t_s$   to zero  at large time. ( Right) $N_{bulk}$ is
increasing  from zero  at $t=t_s$ large value for large time. We
have assumed $p=3$ for $3+1$ dimensional M3 which our universe is
located on it and time of collision between branes $t_s=33 Gyr $.}
\end{figure*}


\section{Summary and Discussion} \label{sum}
In this paper, we have investigated the Padmanhabhan proposal in
a system of oscillating branes. In this model, first, a pair of
$M1$-anti-$M1$-branes are constructed from joining $M0$-branes.
During the processes of formation of these branes, two types of
fields emerge. The first type is a  scalar field which moves
in transverse direction and when glues to branes, plays the role
of a graviton scalar mode. The second type lives on the brane,
plays the role of graviton tensor modes  and causes the formation of
a wormhole  between the branes. By closing two branes towards
each other, the wormhole dissolves in them and leads to an
inequality between the number of degrees of freedom on the  surface of the
branes and in the bulk. Near
the colliding point, the square of energy of system becomes
negative and for solving this problem, the $M1$-branes compact, two-form gauge fields convert to one-form gravity with opposite sign
and anti-gravity comes out. In these conditions, branes get away
from each other and their size decreases. By joining $M1$-branes,
higher dimensional branes like $M3$-branes are produced which
compact and open like the initial $M1$'s. Our universe is located
on one of these $M3$-branes and by compacting them, contracts and
by opening, expands. By expanding universe, the number of degrees
of freedom on the surface increases, while the one in the bulk
decreases. However, by contracting universe, the number of degrees of freedom
on the surface decreases and the one in bulk enhances.
In a forthcoming paper, possible observational signatures of this dynamics will be discussed.

\section*{Acknowledgments}
\noindent The work of Alireza Sepehri has been supported
financially by Research Institute for Astronomy and Astrophysics
of Maragha (RIAAM),Iran under research project No.1/4165-14. SC
acknowledges the Tomsk State Pedagogical University (TSPU) for
having awarded the title of Honorary Professor. SC is supported by
INFN ({\it iniziative specifiche} QGSKY and TEONGRAV).

 \end{document}